# Curvature and frontier orbital energies in density functional theory

Tamar Stein[1], Jochen Autschbach,[2] Niranjan Govind,[3] Leeor Kronik*[4], and Roi Baer*[1]

[1]*Fritz Haber center for Molecular Dynamics, The Institute of Chemistry, The Hebrew University, Jerusalem 91904, Israel.*
[2]*Department of Chemistry, University at Buffalo, Buffalo, NY 14260, USA*
[3]*William R. Wiley Environmental Molecular Sciences Laboratory, Pacific Northwest National Laboratory, Richland, WA 99352, USA*
[4]*Department of Materials and Interfaces Weizmann Institute of Science, Rehovoth 76100, Israel.*

Perdew *et al.* [Phys. Rev. Lett. **49**, 1691 (1982)] discovered and proved two different properties of exact Kohn-Sham density functional theory (DFT): (i) The exact total energy versus particle number is a series of linear segments between integer electron points; (ii) Across an integer number of electrons, the exchange-correlation potential may ``jump'' by a constant, known as the derivative discontinuity (DD). Here, we show analytically that in both the original and the generalized Kohn-Sham formulation of DFT, the two are in fact two sides of the same coin. Absence of a derivative discontinuity necessitates deviation from piece-wise linearity, and the latter can be used to correct for the former, thereby restoring the physical meaning of the orbital energies. Using selected small molecules, we show that this results in a simple correction scheme for any underlying functional, including semi-local and hybrid functionals as well as Hartree-Fock theory, suggesting a practical correction for the infamous gap problem of DFT. Moreover, we show that optimally-tuned range-separated hybrid functionals can inherently minimize *both* DD and curvature, thus requiring no correction, and show that this can be used as a sound theoretical basis for novel tuning strategies.

In a seminal article,[1] Perdew, Parr, Levy, and Balduz (PPLB) discovered and proved two properties that exact density functional theory (DFT) must obey. By extending the realm of DFT to fractional electron numbers, using a zero temperature statistical mixture of integer electron states, they found that: (i) The exact total energy versus particle number curve must be a series of linear segments between the integer electron points; (ii) Across an integer number of electrons, the exchange-correlation (XC) potential may ``jump'' by a constant, usually known as the derivative discontinuity (DD) $\Delta_{XC}$.[2] Originally derived to explain how DFT handles the dissociation of heterodimers, both properties were later found important in a host of other contexts. Piecewise linearity and its generalization to spin densities were deemed essential to avoiding localization/delocalization and static correlation errors[3] and the absence of DD in approximate KS functionals have also been implicated in erroneous description of charge transfer processes.[4,5]

The DD has been found to play a decisive and negative role in the relation between the frontier, highest occupied (HOMO) and lowest unoccupied (LUMO) Kohn-Sham (KS) orbital energies, (OEs) and the ionization potential (IP) and electron affinity (EA), respectively.[6] In principle, with the exact (asymptotically vanishing) KS potential the HOMO energy can be identified with the IP.[1,7-11] Potentials derived from approximate functionals, that are analytical and therefore do not possess a DD, fail to achieve this. Worse, unless this "missing" DD (MDD) is negligible the IP and the EA cannot be *simultaneously* identified with the HOMO and LUMO energies even with the exact KS potential, leading to the infamous "gap problem"[12,13]. These conclusions are equally valid for generalized KS (GKS) theory[14,15], where a non-local potential operator may be present[16].

In this article, we show that the MDD and the deviation from piecewise linearity are in fact doppelgängers.[17] The former *necessitates* the latter, and the latter can be used to correct for the former and restore the physical meaning of the frontier OEs. We further show that this results in a simple and robust procedure for quantitative correction of these OEs for any underlying KS or generalized Kohn-Sham (GKS) DFT. Specifically, we show that optimally-tuned range-separated hybrid (RSH) functionals minimizes both MDD and curvature and that this can be used as a basis for novel tuning strategies.

To set notation, we consider Coulomb systems with an average number of electrons equal to $M = N + x$, where $N$ is an integer and $-1 \leq x \leq 0$ is a "fractional hole". Denoting the total energy of such a system by $E^N(x)$, piecewise linearity dictates that

$$E^N(x) = (1 + x)E^N - xE^{N-1}, \quad (1)$$

such that the energy of the $N$ and $N-1$ electron systems is given by $E^N = E^N(0)$ and $E^{N-1} = E^N(-1)$, respectively, with the (vertical) IP of the $N$ electron system defined as $I^N \equiv E^{N-1} - E^N$. From Eq. (1), $-I^N$ is the constant slope of $E_N(x)$ in the range $-1 \leq x \leq 0$. Similar considerations show that as the electron number changes through $N$, from $N + x$ to $N - x$, the slope changes discontinuously, with the new slope given by $-I^{N+1}$.[1,15,18] The latter is, by definition, the EA of the $N$ electron system.

Exact KS-DFT maps the fractional charge interacting-electron ensemble onto a non-interacting system with a fractionally-occupied HOMO, possessing the same electron density. Remarkably, the energy of the fractionally-occupied HOMO, $\varepsilon_H^N(x)$, must remain constant as a function of $x$.[19] This is a consequence of Janak's theorem[20], according to which $\varepsilon_H^N(x) = \frac{d}{dx}E^N(x)$, which is constant equal to $-I^N$. As $M$ changes from slightly below $N$ to slightly above $N$, a new orbital is fractionally occupied, such that $\varepsilon_H^{N+1} = -I^{N+1}$. PPLB introduced the DD to argue that $-I^{N+1}$ will usually deviate from $\varepsilon_L^N(x \to 0^-)$ because the KS potential ``jumps'' by $\Delta_{XC}$. Thus, at integer $N$, the KS frontier OEs relate to the ionization energies (IEs) as[12,21]:

$$I^N + \varepsilon_H^N = 0 \; ; \; I^{N+1} + \varepsilon_L^N = -\Delta_{XC} \quad (2)$$

and consequently the KS gap, $\varepsilon_L^N - \varepsilon_H^N$, is related to the fundamental gap, $I^N - I^{N+1}$, by

$$\varepsilon_L^N - \varepsilon_H^N + \Delta_{XC} = I^N - I^{N+1} \quad (3)$$

* [roi.baer@huji.ac.il](mailto:roi.baer@huji.ac.il) ; [leeor.kronik@weizmann.ac.il](mailto:leeor.kronik@weizmann.ac.il)

Numerical estimates for many systems have shown that $\Delta_{XC}$ can be quite large.[22, 23] For bulk Si it is ~0.5 eV, which is nearly 50% of the fundamental gap[24].

*Approximate* KS functionals, notably local/semi-local functionals used in most DFT applications, typically employ analytical, explicitly density-dependent XC functionals that do not allow for a DD. Almost without exception, it is found numerically that $\varepsilon_H^N$ and $\varepsilon_L^N$ obtained from such functionals respectively underestimate and overestimate by similar amounts the IP and EA computed from total energy differences using the same functional[21]. Explicitly:

$$I^N + \varepsilon_H^N \approx \frac{1}{2}\delta_{XC} \ ; \ I^{N+1} + \varepsilon_L^N \approx -\frac{1}{2}\delta_{XC} \quad (4)$$

where $\delta_{XC}$ denotes the deviation between the fundamental and KS gaps :

$$\varepsilon_L^N - \varepsilon_H^N + \delta_{XC} \approx I^N - I^{N+1}, \quad (5)$$

One should compare these relations in these *approximate*, DD-free KS functionals with Eqs. (2) and (3) valid in *exact* KS DFT. Consider now that energies from approximate functionals exhibit non-zero curvature, $C^N(x) \equiv \frac{d^2}{dx^2}E^N(x) \neq 0$. Let us first approximate $C^N(x)$ as a constant, $C^N$. The ensemble energy within this (G)KS approximation then deviates from Eq. (1), given by:

$$E^N(x) = \frac{1}{2}C^N x(x+1) + (1+x)E^N - xE^{N-1}. \quad (6)$$

Janak's theorem,[20] allows expression of the OEs by derivatives of the energies:

$$\varepsilon_H^N = \frac{dE^N(x)}{dx}\bigg|_{x=0}, \quad \varepsilon_L^{N-1} = \frac{dE^N(x)}{dx}\bigg|_{x=-1}, \quad (7)$$

where for the second equality we used the fact that for a DD-free functional $\varepsilon_L(x \to -1^-) = \varepsilon_H(x \to -1^+)$. Inserting Eq. (6) in (7) yields $I^N + \varepsilon_H^N = \frac{1}{2}C^N$ and $I^{N+1} + \varepsilon_L^N = -\frac{1}{2}C^{N+1}$.

For perfectly parabolic $E(x)$ curves, then, half the curvature *exactly* corrects for the difference between OEs and IEs. Consequently, the average of the two curvatures exactly corrects the gap: $I^N - I^{N+1} = \varepsilon_L^N - \varepsilon_H^N + \frac{1}{2}(C^N + C^{N+1})$. Generally, $C^N(x)$ does depend on $x$, but a similar expansion shows that to leading order

$$I^N \approx I^N(\text{H}) \equiv -\varepsilon_H^N + \frac{1}{2}C^N(0)$$
$$I^{N+1} \approx I^{N+1}(\text{L}) \equiv -\varepsilon_L^N - \frac{1}{2}C^{N+1}(-1) \quad (8)$$
$$I^N - I^{N+1} \approx \varepsilon_L^N - \varepsilon_H^N + \frac{1}{2}(C^N + C^{N+1})$$

with the additional exact relation:

$$\bar{C}^N \equiv \int_{-1}^{0} C^N(x)dx = \varepsilon_H^N - \varepsilon_L^{N-1}, \quad (9)$$

where the last equality in the above equation follows from Eq. (7).

Eqs. (8) serve as a natural basis for a straightforward "curvature correction" to the OEs of approximate, DD-free density functionals. To study its efficacy, we applied it to PBE calculations of selected small molecules. As shown in Table I, the negative of the neutral HOMO energy, $-\varepsilon_H^N$, and of the cation LUMO energy, $-\varepsilon_L^{N-1}$, both deviate by an average of more than 4 eV from the neutral IP obtained from total energy differences, $I^N$. However, the curvature values $C^N(0)$ and $C^N(-1)$, also given in Table 1, are large numbers, spanning a wide range of values from 4 eV to 13 eV. The curvature-corrected estimates, $I^N(H)$ and $I^N(L)$ of Eqs. (8), are remarkably close to $I^N$, with absolute mean deviation of less than ~0.1eV.

For simplicity, we have focused our discussion so far on KS theory. However, as mentioned above, piecewise linearity must also be obeyed by GKS theory, i.e., even in the presence of a non-local potential.[14, 15] The local "remainder potential" of GKS theory [16], which plays a role analogous to that of the exchange-correlation potential in KS theory, can similarly exhibit a DD [14]. Therefore, Eqs. (8) are immediately valid for any GKS functional whose "remainder potential" *does not* possess a DD. The curvature-based correction scheme should therefore apply to any hybrid functional as well, as such functionals are merely a special case of GKS theory[6, 14, 25, 26].

Table 1 further shows that our correction procedure is indeed equally helpful for the PBE-based hybrid, PBE0.[27-29] Here curvatures are somewhat smaller than for PBE and OEs are closer to $I_N$, but the deviation is still large, close to ~3 eV and after correction ~0.1 eV. Furthermore, the same correction assists Hartree-Fock (HF) theory (HFT), itself just another special GKS case.[6, 16] Indeed, the HF uncorrected gaps in Table 1, are too large and curvatures are negative (implying a *negative* MDD) while curvature-corrected IEs are significantly closer to $I_N$ with mean deviance of ~0.3 eV. This is larger due to enlarged non-parabolic behavior of $E(x)$ seen in Table 1 as a larger mean difference between $C^N(0)$ and $C^N(-1)$.

Further insight can be attained from the following physical arguments. By using an approximate functional devoid of a DD, we rob the (G)KS potential of its ability to correctly follow the slope change of the ideally piecewise-linear energy versus fractional electron number curve. Concomitantly, for the calculation to be useful at all, we still expect the approximate density functional to deliver rather accurate energy values at the integer electron points. Not unlike nudging an elastic band nailed at regular intervals, this is possible only by introducing curvature. Hence, *compromising on the DD inevitably necessitates curvature.* Conversely, precisely because the curvature must compensate for the MDD, *curvature corrections restore the physical meaning of the (G)KS frontier eigenvalues.* Thus, MDD and curvature deviations are inexorably linked. In fact, Eq. (9) shows that the difference between the HOMO eigenvalue of the $N$ electron system and the LUMO value of the $N-1$ electron system, which in the exact theory is precisely the DD[1, 11, 13], is *exactly* the average curvature in the approximate theory. Eqs. (8) fully rationalize the general similarity between $\Delta_{XC}$ of Eqs. (2) and $\delta_{XC}$ of Eqs. (4) of the exact and approximate theory, respectively, with the approximate (G)KS gap mimicking the exact one as long as $C_N \approx C_{N+1}$.

The doppelgänger[17] nature of the MDD and the curvature suggests that their co-appearance also implies their possible co-disappearance: a functional with small MDD will necessarily exhibit small curvature. Recently, we have shown that the MDD can indeed be rigorously minimized, allowing for direct identification of HOMO and LUMO eigenvalues with the IP and EA. This has been accomplished within a GKS scheme, based on a range-split hybrid functional (RSH)[30] that uses exact non-local exchange only for the long-range part of the Coulomb repulsion, $\text{erf}(\gamma r)/r$. The range-splitting parameter $\gamma$ was tuned from first principles, per system, based on satisfaction of the IP ("Koopmans") theorem (i.e., the first of Eqs. (2)),[31, 32] as best as possible, for both the neutral and anionic system [14, 33-35]. In Table 2, we show for selected small molecules that tuning $\gamma$ based on (average or endpoint) curvature minimization yields optimal $\gamma$ values that change by very little compared to those obtained from satisfying the IP theorem for the neutral molecule. Furthermore, the optimal $\gamma$ found from one tuning criterion also fulfills the other two criteria to within ~0.05 eV, i.e., to within the above-obtained accuracy of curvature corrections.

Table 1: Calculated PBE, PBE0, and HF OEs $\varepsilon_H^N$ and $\varepsilon_L^{N-1}$, endpoint curvatures $C^N(x)$ ($x = 0, -1$), average curvature $\bar{C}^N = \epsilon_H^N - \epsilon_L^{N-1}$, curvature-corrected estimates $I^N(H)$ and $I^N(L)$, and IP obtained from total energy differences, $I^N$, for a set of selected molecules. All values are in eV. All calculations were performed with a development version of NWChem,[36] using the cc-pVTZ basis and B3LYP/cc-pVTZ optimized molecular geometries.

| Functional | | $C^N(0)$ | $C^N(-1)$ | $\bar{C}^N$ | $-\varepsilon_H^N$ | $-\varepsilon_L^{N-1}$ | $I^N(H)$ | $I^N(L)$ | $I^N$ |
|---|---|---|---|---|---|---|---|---|---|
| PBE | $F_2$ | 12.0 | 12.5 | 12.3 | 9.1 | 21.4 | 15.1 | 15.1 | 15.2 |
| | $N_2$ | 10.2 | 10.5 | 10.4 | 10.2 | 20.6 | 15.3 | 15.4 | 15.4 |
| | $NH_3$ | 10.0 | 8.8 | 9.7 | 5.9 | 15.6 | 10.9 | 11.2 | 10.9 |
| | $H_2O$ | 11.6 | 10.7 | 11.5 | 6.8 | 18.3 | 12.6 | 13.0 | 12.6 |
| | $CH_2O$ | 9.0 | 9.1 | 9.2 | 6.1 | 15.3 | 10.6 | 10.7 | 10.7 |
| | HCOOH | 8.9 | 8.7 | 8.9 | 6.8 | 15.7 | 11.2 | 11.3 | 11.2 |
| | Benzene | 6.2 | 5.9 | 6.0 | 6.3 | 12.2 | 9.4 | 9.3 | 9.3 |
| | Naphthalene | 4.9 | 4.9 | 4.9 | 5.5 | 10.4 | 7.9 | 7.9 | 7.9 |
| | Anthracene | 4.3 | 4.2 | 4.3 | 5.0 | 9.2 | 7.1 | 7.1 | 7.1 |
| | Mean Deviation from $I_N$ | | | | -4.3 | 4.3 | -0.02 | 0.08 | |
| | Mean Absolute Deviation from $I_N$ | | | | 4.3 | 4.3 | 0.05 | 0.10 | |
| PBE0 | $F_2$ | 8.2 | 8.6 | 8.4 | 11.6 | 20.0 | 15.7 | 15.7 | 15.8 |
| | $N_2$ | 7.1 | 7.2 | 7.3 | 12.1 | 19.4 | 15.7 | 15.8 | 15.7 |
| | $NH_3$ | 6.5 | 5.6 | 6.3 | 7.5 | 13.8 | 10.8 | 11.0 | 10.8 |
| | $H_2O$ | 7.4 | 6.8 | 7.3 | 8.8 | 16.1 | 12.5 | 12.7 | 12.5 |
| | $CH_2O$ | 5.8 | 5.8 | 5.9 | 7.8 | 13.7 | 10.7 | 10.8 | 10.7 |
| | HCOOH | 5.5 | 5.4 | 5.6 | 8.5 | 14.1 | 11.3 | 11.4 | 11.3 |
| | Benzene | 4.0 | 3.9 | 4.0 | 7.3 | 11.2 | 9.3 | 9.3 | 9.3 |
| | Naphthalene | 3.3 | 3.3 | 3.3 | 6.3 | 9.6 | 8.0 | 8.0 | 8.0 |
| | Anthracene | 2.9 | 2.9 | 2.9 | 5.7 | 8.6 | 7.2 | 7.2 | 7.2 |
| | Mean Deviation from $I_N$ | | | | -2.8 | 2.8 | -0.01 | 0.07 | |
| | Mean Absolute Deviation from $I_N$ | | | | 2.8 | 2.8 | 0.01 | 0.09 | |
| HF | $F_2$ | -4.3 | -3.1 | -3.7 | 18.0 | 14.3 | 15.9 | 15.9 | 16.1 |
| | $N_2$ | -3.3 | -2.7 | -2.6 | 16.8 | 14.2 | 15.1 | 15.5 | 15.7 |
| | $NH_3$ | -5.8 | -2.8 | -4.0 | 11.6 | 7.6 | 8.7 | 9.0 | 9.3 |
| | $H_2O$ | -7.2 | -3.5 | -5.0 | 13.7 | 8.7 | 10.1 | 10.4 | 10.9 |
| | $CH_2O$ | -5.4 | -3.9 | -4.9 | 12.0 | 7.1 | 9.3 | 9.0 | 9.4 |
| | HCOOH | -6.3 | -3.8 | -5.3 | 12.9 | 7.6 | 9.7 | 9.5 | 10.0 |
| | Benzene | -2.6 | -1.9 | -2.3 | 9.1 | 6.9 | 7.8 | 7.9 | 8.0 |
| | Naphthalene | -2.0 | -1.8 | -1.9 | 7.9 | 6.0 | 6.9 | 6.2 | 6.9 |
| | Anthracene | -1.8 | -1.8 | -1.8 | 7.0 | 5.2 | 6.1 | 6.1 | 6.1 |
| | Mean Deviation from $I_N$ | | | | 1.9 | -1.6 | -0.3 | -0.3 | |
| | Mean Absolute Deviation from $I_N$ | | | | 1.8 | 1.6 | 0.3 | 0.3 | |

The above argumentation immediately explains the hitherto heuristic observation that optimally-tuned RSHs tend to have small curvature [14, 37, 38] and provides a solid theoretical framework for more elaborate RSHs, where both range-separation parameter and short-range exact-exchange fraction are tuned by demanding minimization of both MDD and curvature [39, 40].

We note that tuning using endpoint curvature is additionally advantageous in that it may be performed entirely on the original system, without recourse to its ionized or even partially ionized states - a major advantage for larger systems, or for solids where ionization is more challenging owing to periodic boundary conditions. This can be accomplished by considering that by combining Janak's theorem [Eq. (7)] with the curvature definition, we find that requisite curvatures are given by $C_N = \frac{\partial \epsilon_H(N)}{\partial f_H}$ and $C_{N+1} = \frac{\partial \epsilon_L(N)}{\partial f_L}$. Treating orbital occupation as a perturbation, we in the supplementary material a linear-response equation[32] whose solution includes the curvatures $\partial \epsilon_n / \partial f_n$ with respect to any KS orbital $\psi_n$. Starting from an analog of the single-pole approximation in linear-response time-dependent DFT[41] we also derive an order-by-order expansion for the curvature. The zero order estimate for the curvature (see also ref. 32) equals to the OE correction estimate of ref. 42. However, we also demonstrate, using $H_2O$, that this bare "self-interaction" term grossly overestimates curvature because of extensive relaxation entering in higher orders. In the supplementary material, we show that one needs to go beyond even 2nd order when estimating curvature.

In recent years, several orbital-dependent corrections of (semi-)local functionals have been put forth (e.g., refs. [43-47]). These provide a route different from ours for overcoming the DD-curvature equivalence by formally departing from the (G)KS framework because for physically meaningful results, any curvature minimization must be accompanied by a mechanism of MDD reduction.

Table 2: The optimal value of the range parameter $\gamma$ (in atomic units) in the BNL functional,[31] as determined by three different first principles tuning criteria.

| | Koopmans': $\varepsilon_H^N = -I^N$ | Zero average curvature: $\varepsilon_H^N = \varepsilon_L^{N-1}$ | Zero endpoint curvature: $C^N = 0$ |
|---|---|---|---|
| $F_2$ | 0.73 | 0.74 | 0.71 |
| $N_2$ | 0.61 | 0.61 | 0.59 |
| $NH_3$ | 0.50 | 0.51 | 0.48 |
| $H_2O$ | 0.57 | 0.58 | 0.55 |
| $CH_2O$ | 0.49 | 0.49 | 0.48 |
| HCOOH | 0.46 | 0.46 | 0.45 |
| Benzene | 0.31 | 0.32 | 0.31 |
| Naphthalene | 0.28 | 0.28 | 0.28 |
| Anthracene | 0.25 | 0.25 | 0.25 |

A prevalent view is that self-interaction errors cause curvature and therefore self-interaction corrections remove it. Yet, it was already noticed that functionals that are formally one- and two-self-interaction free can still deviate from piecewise linearity, [19,49] leading to the suggestion to identify curvature with "many-electron self-interaction errors (SIEs)". Our approach is to interpret curvature as required by MDD. Self-interaction is then just a mechanism for creating the required curvature. Other mechanisms are possible. Indeed, curvature can arise in functionals that are formally self-interaction free such as in HFT.

In conclusion, we have shown analytically that absence of a DD and deviation from piecewise linearity are, quantitatively, two sides of the same coin. We have used this to show how curvature can be used to estimate quite accurately the MDD and to estimate the deviation of the frontier OEs from the IP and EA. As a special case, we showed that optimally-tuned RSHs designed to minimize the DD and bring OEs close to the corresponding IPs must also mitigate curvature. This allowed us to suggest the minimization of curvature obtained from a linear-response-like formalism as a tuning criterion and to examine its relation to single-electron self-interaction errors. Importantly, all claims were established quantitatively using a set of small molecules and selected semi-local, hybrid, RSH, and HF calculations.

Work in Jerusalem was supported by the Israel Science Foundation and in Rehovoth by the European Research Council and the Lise Meitner Minerva Center for Computational Chemistry. Development of the fractional-occupation code in NWChem was supported by DOE grant DE-FG02-09ER16066 (BES, Heavy Element Chemistry) to JA.

[1] J. P. Perdew, R. G. Parr, M. Levy, and J. L. Balduz, Phys. Rev. Lett. **49**, 1691 (1982).
[2] These two theorems and all results presented in this article refer to zero temperature ensembles. Extension to finite-temperature ensembles is discussed in ref. 1, with some comments in the supplementary material to this article.
[3] A. J. Cohen, P. Mori-Sanchez, and W. T. Yang, Science **321**, 792 (2008).
[4] D. J. Tozer, J. Chem. Phys. **119**, 12697 (2003).
[5] S. Kümmel, L. Kronik, and J. P. Perdew, Phys. Rev. Lett. **93**, 213002 (2004).
[6] S. Kümmel and L. Kronik, Rev. Mod. Phys. **80**, 3 (2008).
[7] C.-O. Almbladh and U. von-Barth, Phys. Rev. B **31**, 3231 (1985).
[8] J. P. Perdew and M. Levy, Phys. Rev. B **56**, 16021 (1997).
[9] M. Levy, J. P. Perdew, and V. Sahni, Phys. Rev. A **30**, 2745 (1984).
[10] D. P. Chong, O. V. Gritsenko, and E. J. Baerends, J. Chem. Phys. **116**, 1760 (2002).
[11] O. V. Gritsenko and E. J. Baerends, J. Chem. Phys. **117**, 9154 (2002).
[12] J. P. Perdew and M. Levy, Phys. Rev. Lett. **51**, 1884 (1983).
[13] L. J. Sham and M. Schlüter, Phys. Rev. Lett. **51**, 1888 (1983).
[14] L. Kronik, T. Stein, S. Refaely-Abramson, and R. Baer, J. Chem. Theor. Comp **8**, 1515 (2012).
[15] W. Yang, A. J. Cohen, and P. Mori-Sanchez, J. Chem. Phys. **136**, 204111 (2012).
[16] A. Seidl, A. Görling, P. Vogl, J. A. Majewski, and M. Levy, Phys. Rev. B **53**, 3764 (1996).
[17] This term was originally introduced by J. P. Perdew [Adv. Chem. Phys. 21, 113 (1990)] to emphasize links between the DD and self-interaction corrections. Here we establish quantitative relations between the DD and piecewise linearity.
[18] A. J. Cohen, P. Mori-Sanchez, and W. T. Yang, Phys. Rev. B **77**, 115123 (2008).
[19] A. Ruzsinszky, J. P. Perdew, G. I. Csonka, O. A. Vydrov, and G. E. Scuseria, J. Chem. Phys. **126**, 104102 (2007).
[20] J. Janak, Phys. Rev. B **18**, 7165 (1978).
[21] A. M. Teale, F. De Proft, and D. J. Tozer, J. Chem. Phys. **129**, 044110 (2008).
[22] G. K.-L. Chan, J. Chem. Phys. **110**, 4710 (1999).
[23] M. J. Allen and D. J. Tozer, Mol. Phys. **100**, 433 (2002).
[24] R. W. Godby, M. Schluter, and L. J. Sham, Phys. Rev. Lett. **56**, 2415 (1986).
[25] A. Görling and M. Levy, J. Chem. Phys. **106**, 2675 (1997).
[26] R. Baer, E. Livshits, and U. Salzner, Ann. Rev. Phys. Chem. **61**, 85 (2010).
[27] J. P. Perdew, K. Burke, and M. Ernzerhof, Phys. Rev. Lett. **77**, 3865 (1996).
[28] C. Adamo and V. Barone, J. Chem. Phys. **110**, 6158 (1999).
[29] M. Ernzerhof and G. E. Scuseria, J. Chem. Phys. **110**, 5029 (1999).
[30] A. Savin, in *Recent Advances in Density Functional Methods Part I*, edited by D. P. Chong (World Scientific, Singapore, 1995), p. 129.
[31] E. Livshits and R. Baer, Phys. Chem. Chem. Phys. **9**, 2932 (2007).
[32] U. Salzner and R. Baer, J. Chem. Phys. **131**, 231101 (2009).
[33] T. Stein, H. Eisenberg, L. Kronik, and R. Baer, Phys. Rev. Lett. **105**, 266802 (2010).
[34] S. Refaely-Abramson, R. Baer, and L. Kronik, Phys. Rev. B **84**, 075144 (2011).
[35] T. Stein, L. Kronik, and R. Baer, J. Chem. Phys. **131**, 244119 (2009).
[36] M. Valiev, et al., Comp. Phys. Comm. **181**, 1477 (2010).
[37] M. Srebro and J. Autschbach, J. Chem. Theor. Comp **8**, 245 (2011).
[38] M. Srebro and J. Autschbach, J. Phys. Chem. Lett. **3**, 576 (2012).
[39] M. Srebro, N. Govind, W. A. de Jong, and J. Autschbach, J. Phys. Chem. A **115**, 10930 (2011).
[40] S. Refaely-Abramson, R. Baer, and L. Kronik, ArXiv.org (2012).
[41] M. Petersilka, U. J. Gossmann, and E. K. U. Gross, Phys. Rev. Lett. **76**, 1212 (1996).
[42] T. Ziegler, M. Seth, M. Krykunov, J. Autschbach, and F. Wang, Theochem-J. Mol. Struct. **914**, 106 (2009).
[43] M. Cococcioni and S. de Gironcoli, Phys. Rev. B **71**, 035105 (2005).
[44] S. Lany and A. Zunger, Phys. Rev. B **81**, 205209 (2010).
[45] I. Dabo, A. Ferretti, N. Poilvert, Y. L. Li, N. Marzari, and M. Cococcioni, Phys. Rev. B **82**, 115121 (2010).
[46] X. Zheng, A. J. Cohen, P. Mori-Sánchez, X. Hu, and W. Yang, Phys. Rev. Lett. **107**, 026403 (2011).
[47] J. Jellinek and P. H. Acioli, J. Chem. Phys. **118**, 7783 (2003).